\pdfoutput=1 		

\documentclass[letterpaper, 10 pt, conference]{ieeeconf}  

\IEEEoverridecommandlockouts                              
\usepackage[cmex10]{amsmath}
\usepackage{amsfonts,amssymb,bm}

\usepackage{graphicx,color,psfrag}		
\usepackage{cite}			
\usepackage{subfigure}
\usepackage[hidelinks]{hyperref}
\let\labelindent\relax 		
\usepackage{enumitem}

\usepackage{epstopdf}	

\graphicspath{{.},{figures/}}
\newcommand{\transpose}{\text{T}}
\newcommand{\transp}{\transpose}

\newcommand{\Yall}{\mathcal{Y}}
\newcommand{\Gamall}{\Gamma}
\newcommand{\xKF}{\hat{x}^\text{F}}
\newcommand{\PKF}{P^\text{F}}
\newcommand{\PKFss}{\bar{P}^\text{F}}
\newcommand{\eKF}{e^\text{F}}
\newcommand{\xI}{\hat{x}^\text{I}}
\newcommand{\xII}{\hat{x}^\text{II}}
\newcommand{\eI}{e^{\text{I}}}
\newcommand{\eII}{e^{\text{II}}}
\newcommand{\eIhat}{\hat{e}^{\text{I}}}
\newcommand{\eIIhat}{\hat{e}^{\text{II}}}
\newcommand{\PI}{P^{\text{I}}}
\newcommand{\PII}{P^{\text{II}}}

\newcommand{\Vo}{V_\text{o}}
\newcommand{\Vok}[1]{V_{\text{o},#1}}

\newcommand{\commC}{C} 	
\newcommand{\last}{\ell}			
\newcommand{\lastel}{\kappa}		

\newcommand{\trigsig}{\bar{E}}
\newcommand{\trigsigM}{\bar{E}^\text{mean}}
\newcommand{\trigsigV}{\bar{E}^\text{var}}

\newcommand{\graph}[1]{{\bf #1}}

\providecommand{\norm}[1]{\lVert#1\rVert}

\newcommand{\Cc}{\mathcal{C}}
\newcommand{\Dc}{\mathcal{D}}
\newcommand{\Ec}{\mathcal{E}}

\newcommand{\Nc}{\mathcal{N}}

\newcommand{\Yc}{\mathcal{Y}}

\newcommand{\field}[1]{\mathbb{#1}}
\newcommand{\R}{\field{R}}

\DeclareMathOperator*{\E}{\field{E}}
\DeclareMathOperator*{\Var}{Var}

\newcommand{\trace}{\text{trace}}

\newtheorem{lemma}{Lemma}

\newtheorem{example}{Example}
\newtheorem{remark}{Remark}

\newcommand{\ie}{i\/.\/e\/.,\/~}
\newcommand{\eg}{e\/.\/g\/.,\/~}
\newcommand{\cf}{cf\/.\/~}
\newcommand{\fig}{Fig\/.\/~}

\newcommand{\sect}{Sec\/.\/~}

\newcommand{\Lem}{Lemma~}

\newcommand{\Ex}{Example~}

\title{\LARGE \bf
Predictive and Self Triggering for Event-based State Estimation
}

\author{Sebastian Trimpe
\thanks{S. Trimpe is with the Autonomous Motion Department at the Max Planck Institute for Intelligent Systems, 
	72076 T\"ubingen, Germany. 
	\href{mailto:strimpe@tuebingen.mpg.de}{\tt strimpe@tuebingen.mpg.de}}%
\thanks{This work was supported in part by the Max Planck Society, the Max Planck ETH Center for Learning Systems, and the German Research Foundation (DFG) within the Priority Program 1914.}
}

\usepackage{fancyhdr}
\newcommand{\mytitle}{\textbf{Accepted final version.}
To appear in \textit{Proc. of the 55th IEEE Conference on Decision and Control, 2016}.\\
\copyright 2016 IEEE. Personal use of this material is permitted. Permission from IEEE must be obtained for all other uses, in any current or future media, including reprinting/republishing this material for advertising or promotional purposes, creating new collective works, for resale or redistribution to servers or lists, or reuse of any copyrighted component of this work in other works.}
\begin{document}

\maketitle

\thispagestyle{fancy}	
\pagestyle{empty}

\begin{abstract}
Event-based state estimation can achieve estimation quality comparable to traditional time-triggered methods, but with a significantly lower number of samples. In networked estimation problems, this reduction in sampling instants does, however, not necessarily translate into better usage of the shared communication resource. Because typical event-based approaches decide instantaneously whether communication is needed or not, free slots cannot be reallocated immediately, and hence remain unused. In this paper, novel predictive and self triggering protocols are proposed, which give the communication system time to adapt and reallocate freed resources. From a unified Bayesian decision framework, two schemes are developed: self-triggers that predict, at the current triggering instant, the next one; and predictive triggers that indicate, at every time step, whether communication will be needed at a given prediction horizon. The effectiveness of the proposed triggers in trading off estimation quality for communication reduction is compared in numerical simulations.

\end{abstract}

\section{Introduction}
\label{sec:intro}
In recent years, the research community in event-based control and state estimation has had remarkable success in showing
that the number of samples in feedback loops can be reduced significantly, as compared to traditional time-triggered designs.  
The resulting reduction in average communication or processing can be translated into increased battery life \cite{ArMaAnTaJo13} in wireless sensor systems, for example.
However, it has rarely been demonstrated that event-based designs also
result in
better utilization of shared communication and processing resources, or reduced hardware costs.  

A fundamental problem of most event-triggered designs is that they make decisions about whether a communication or control computation is needed \emph{instantaneously}.
This means that the resource must be held available at all times in case of a positive triggering decision.  Conversely, if a triggering decision is negative, the reserved slot 
remains unused because it cannot be reallocated to other users immediately.

In order to translate the reduction in average sampling rates 
to better \emph{actual} resource utilization, it is vital that the event-based system is able to \emph{predict} resource usage ahead of time, rather than requesting resources instantaneously.  
This allows the processing or communication system  to reconfigure and make unneeded resources available to other users or processes.  Developing such predictive triggering mechanisms for event-based state estimation
is the main objective of this paper.
%
%
\begin{figure}[tb]
\centering
\includegraphics[width=.96\columnwidth]{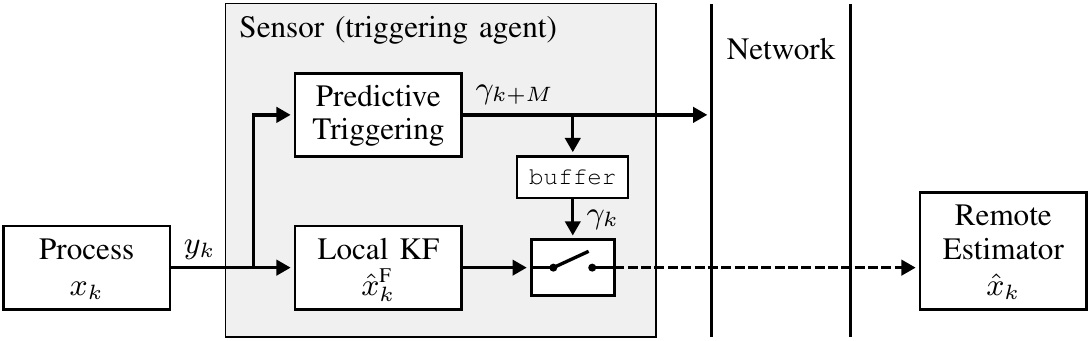}
\caption{Predictive triggering problem.  The sensor runs a local Kalman filter (KF) and transmits its estimate $\xKF_k$ to the remote estimator in case of a positive triggering decision ($\gamma_k = 1$).  The predictive trigger computes the triggering decisions ($\gamma_{k+M} \in \{0,1\}$) $M$ steps ahead of time.  This information can be used by the network to allocate resources.
}
\label{fig:remoteEstimation}
\end{figure}

\subsubsection*{Related work}
The 
area of 
event-based control and estimation has 
substantially grown
in the last decades.  For recent overviews, please refer to \cite{HeJoTa12,GrHiJuEtAl14,Mi15} for control, and to \cite{SiNoLaHa16,ShShCh16} for state estimation, for example.

The concept of \emph{self triggering} has been proposed \cite{VeMaFu03} to address the problem of predicting future sampling instants.  In contrast to event-triggered implementations, which require the continuous monitoring of a triggering signal (such as a control error), self-triggered approaches predict the next triggering instant already at the previous trigger.
Several approaches to self-triggered control have been proposed in literature, see, \eg \cite{HeJoTa12,WaLe09,MaAnTa10,AnTa10,AlSiPa15} and references therein.   
%
Self triggering approaches for state estimation have, however, received less attention.  
Some of the results for estimation are briefly discussed next.

Self triggering is considered for set-valued state estimation in \cite{MePr14}, and for high-gain continuous-discrete observers in \cite{AnNaSeVi15}. In \cite{MePr14}, a new measurement is triggered when the uncertainty set about some part of the state vector becomes too large.  In \cite{AnNaSeVi15}, the triggering rule is designed so as to ensure convergence of the observer.  
The recent works \cite{BrGoHeAl15} and \cite{KoFi15} propose self triggering approaches, where transmission schedules for multiple sensors are optimized at a-priori fixed, periodic time instants taking into account the cost of sampling and estimation/control performance.
While the re-computation of the schedule happens periodically, the transmission of sensor data does generally not.  
In \cite{AlSiPa12}, a discrete-time observer is used as a component of a self-triggered output feedback control system.  Therein, triggering instants are determined by the controller to ensure closed-loop stability.
%
None of the mentioned references considers the approach taken herein, where triggering is formulated as a Bayesian decision problem under different information patterns.

%

%

\subsubsection*{Contributions}
This paper addresses the design of different predictive triggering mechanisms for event-based state estimation.  For this, we consider the remote estimation problem shown in \fig \ref{fig:remoteEstimation}, where a sensor (with sufficient processing capabilities) decides whether and when to communicate its local state estimate to a remote estimator.  
In detail, this paper makes the following contributions:
\begin{itemize}[leftmargin=5mm,label={--}]
\item Extending previous work \cite{TrCa15} on event trigger design,
we propose a unified decision framework for developing different predictive triggering mechanisms, 
where triggering is formulated as an optimization problem solved under different information patterns.  To the best of the author's knowledge, this provides a new perspective on the triggering problem in estimation.
The framework is used to develop the following two triggering concepts.  

\item First, a \emph{self triggering} rule is derived that predicts the next trigger based on the information available at a current triggering instant. 
The self trigger is closely related to the concept of variance-based triggering \cite{TrDAn14b}, albeit this concept has not been used for self triggering before.

\item 
Second, we propose and develop the concept of \emph{predictive triggering}.  In contrast to self triggering, where the next trigger is computed at the last triggering instant, the predictive trigger continuously monitors the sensor measurements, but predicts a communication $M\!>\!0$ steps ahead of time, where the prediction horizon $M$ is a design parameter.
Predictive triggering is a novel concept, which is situated
between the concepts of event triggering and self triggering.

\item The effectiveness of the different triggers in trading off estimation performance for communication is compared in numerical simulations.
\end{itemize}

%

\subsubsection*{Notation}
We use $f(x)$, $f(x|y)$, and $f(x|\Yc)$ to denote, respectively, the probability density functions (PDFs) of the random variable (RV) $x$, of $x$ conditioned on the RV $y$, and of $x$ conditioned on the set of RVs $\Yc$.  When referring to the RV $x$ conditioned on $y$, we also write $(x|y)$.  $\Nc(x; \mu, \Sigma)$ denotes the PDF of a Gaussian RV $x$ with mean $\mu$ and variance $\Sigma$.  
$\E[\cdot]$ denotes the expected value, and $\Var[\cdot]$ the variance.
For functions $g_1$ and $g_2$, $g_2 \circ g_1$ denotes the composition, \ie $(g_2 \circ g_1)(x) = g_2(g_1(x))$.

\section{Problem} 	
\label{sec:problem}
We consider state estimation of a discrete-time, linear system with Gaussian noise
\begin{align}
x_{k} &= A_{k-1} x_{k-1} + v_{k-1} 	\label{eq:sys_x} \\
y_k &= H_k x_k + w_k 			 \label{eq:sys_y}
\end{align}
with time index $k \! \geq \! 1$; $x_k, v_k \in \R^{n_\text{x}}$, $y_k, w_k \in \R^{n_\text{y}}$; mutually independent random variables 
$x_0$, $v_k$, and $w_k$ with PDFs, respectively, $\Nc(x_0; \bar{x}_0,X_0)$, $\Nc(v_{k}; 0,Q_k)$, and $\Nc(w_k; 0,R_k)$;
 and all matrices of corresponding dimensions.  The set of all measurements up to time $k$  is denoted by $\Yall_k := \{ y_1, y_2, \dots, y_k \}.$
For the successive application of $A_k$ for steps 
$k_1$ to $k_2$, we write $\Phi_{k_2:k_1} = A_{k_2} A_{k_2-1} \cdots A_{k_1+1} A_{k_1}$.


\subsection{Local periodic estimator}
The local estimator has access to all measurements $\Yall_k$, see \fig \ref{fig:remoteEstimation}.  The Kalman filter (KF) is the optimal Bayesian estimator in this setting; it recursively computes the exact conditional PDF $f(x_k| \Yall_k)$, \cite{AnMo05}.
The KF recursion is
\begin{align}
\xKF_{k|k-1} &= A_{k-1} \xKF_{k-1} \label{eq:KF1} \\
\PKF_{k|k-1} &= A_{k-1} \PKF_{k-1} A_{k-1}^\transp + Q_{k-1} =: \Vok{k-1}(\PKF_{k-1}) \label{eq:KF2} \\
L_k &= \PKF_{k|k-1} H_k^\transp (H_k \PKF_{k|k-1} H_k^\transp + R_k)^{-1} \label{eq:KF3_Lk} \\
\xKF_{k} &= \xKF_{k|k-1} + L_k(y_k - H_k \xKF_{k|k-1}) \label{eq:KF4} \\
\PKF_{k} &= (I -L_k H_k) \PKF_{k|k-1}.  \label{eq:KF5}
\end{align}
where $f(x_k | \Yall_{k-1}) = \Nc(x_k; \xKF_{k|k-1}, \PKF_{k|k-1})$, $f(x_k | \Yall_{k})$ $= \Nc(x_k; \xKF_{k}, \PKF_{k})$, and the short-hand notation $\xKF_{k} = \xKF_{k|k}$ and $\PKF_{k} = \PKF_{k|k}$ is used for the posterior variables.
The superscript `F' is used to denote the KF with \emph{full data} in distinction to the later event-based estimator.  
In \eqref{eq:KF2}, we introduced the short-hand $\Vok{k-1}$ for the open-loop variance update for later reference.  

We shall also need the $M$-step ahead prediction of the state ($M \geq 0$); that is, $(x_{k+M} | \Yall_k)$, whose PDF is given by (see \cite[p.~111]{AnMo05})
\begin{equation}
f(x_{k+M} | \Yall_k) = \Nc(x_{k+M}; \, \xKF_{k+M|k}, \PKF_{k+M|k})
\label{eq:PDFstatePredM}
\end{equation}
with mean and variance obtained by the open-loop KF iterations \eqref{eq:KF1}, \eqref{eq:KF2}, \ie
\begin{align}
\xKF_{k+M|k} &= \Phi_{(k+M-1):k} \xKF_k \label{eq:KF_meanPred} \\
\PKF_{k+M|k} &= (\Vok{k+M-1} \circ \cdots \circ \Vok{k+1} \circ \Vok{k}) (\PKF_k) . \label{eq:KF_varPred}
\end{align}

\subsection{Remote event-based estimator}
We consider an event-based architecture, where the sensor sporadically communicates its local estimate $\xKF_k$
to the remote estimator, which, at every step $k$, computes its own state estimate $\hat{x}_k$ from the available data.  
Other event-based architectures  are also conceivable, for example, where measurements $y_k$ instead of state estimates are communicated as in \cite{TrDAn11,TrDAn14b,TrCa15}, which can be beneficial for practical considerations (\eg when $n_\text{y} \ll n_\text{x}$) or in distributed architectures.

We denote by $\gamma_k \in \{0, 1\}$ the decision taken by the sensor about whether an update is sent ($\gamma_k = 1$) or not ($\gamma_k = 0$).  For later reference, we denote the set of all triggering decisions until $k$ by $\Gamall_k := \{ \gamma_1, \gamma_2, \dots, \gamma_k \}$.
%
We abstract communication to be ideal, 
without delay and with zero probability of packet loss.

The remote estimator (\cf \fig \ref{fig:remoteEstimation}) uses the following recursion to compute $\hat{x}_k$, its estimate of $x_k$:
\begin{align}
\hat{x}_k &= 
\begin{cases}
A_{k-1} \hat{x}_{k-1} =: \xI_k & \text{if $\gamma_k = 0$} \\
\xKF_k =: \xII_k & \text{if $\gamma_k = 1$} ;
\end{cases}
\label{eq:remoteEst}
\end{align}
that is, at times when no update is received from the sensor, the estimator simply predicts its previous estimate according to the process model \eqref{eq:sys_x}.  The remote estimator thus corresponds to the open-loop prediction of the KF according to \eqref{eq:KF_meanPred}.  Indeed, let $\last_k \leq k$ denote the last time that data was transmitted; then $\hat{x}_k = \xKF_{k|\last_k}$.

\begin{remark}
Under the assumption of perfect communication, the event of not receiving an update ($\gamma_k=0$) also contains information useful for state estimation (also known as \emph{negative information} \cite{SiNoHa13}).  Here, we disregard this information in the interest of a straightforward estimator implementation (see \cite{TrCa15} for a more detailed discussion).  
\end{remark}

For ease of reference and for distinguishing the two paths that the remote estimator \eqref{eq:remoteEst} can take, we introduced the variables $\xI$ and $\xII$, corresponding to the open-loop estimate and closed-loop estimate, respectively.
Furthermore, we introduce the corresponding errors
\begin{align}
\eI_k &:= x_k - \xI_k \label{eq:eI} \\
\eII_k &:= x_k - \xII_k. \label{eq:eII}
\end{align}
The general estimation error, we denote by $e_k := x_k - \hat{x}_k$.


\subsection{Objective}
\label{sec:objective}
The objective of this paper is the development of principled ways for predicting triggering decisions ahead of time. In particular, we shall develop two concepts:
\begin{enumerate}
\item \emph{predictive triggering:} at every step $k$ and for fixed $M\!>\!0$, $\gamma_{k+M}$ is predicted, \ie whether or not communication is needed at $M$ steps in future; and
\item \emph{self triggering:} the next trigger is predicted at the time of the last trigger.
\end{enumerate}


\section{Triggering Framework} 	
\label{sec:framework}
To develop a framework for making predictive triggering decisions, we extend the approach from \cite{TrCa15}, where the triggering decision is formulated as a one-step optimal decision problem. 
While this framework was used in \cite{TrCa15}  to (re-)derive existing and novel event triggers
 (summarized in \sect \ref{sec:eventTrigger}), we extend the framework herein to yield predictive and self triggering mechanisms (\sect \ref{sec:predTrigger} and \ref{sec:selfTrigger}).

\subsection{Decision framework for event triggering}
\label{sec:eventTrigger}
The triggering agent (`Sensor' in \fig \ref{fig:remoteEstimation}) makes a decision 
between using the communication channel (and thus paying a communication cost $C_k$) to improve the remote estimate, or to save communication, but pay a price in terms of a deteriorated estimation performance (captured by a suitable estimation cost $E_k$).  The communication cost $C_k$ is application specific and may be associated with the use of bandwidth or energy, for example.  We assume $C_k$ is known for all times $k$.  The estimation cost $E_k$ is used to measure the discrepancy between the remote estimation errors without update \eqref{eq:eI} and with update \eqref{eq:eII}; that is, 
\begin{equation}
E_k = E(\eI_k, \eII_k)
\label{eq:estCostFunction}
\end{equation}
for a suitable choice of $E$.
For example,  
\begin{equation}
E_k = (\eI_k)^2 - (\eII_k)^2
\label{eq:Ek_squares_scalar}
\end{equation}
was used in \cite{TrCa15} for scalar quantities. This cost measures in terms of quadratic errors how much worse the error without update ($\eI_k$) is, compared to the one  with update ($\eII_k$).

Formally, the triggering decision can then be written as
\begin{equation}
\min_{\gamma_k \in \{0, 1\}}  \gamma_k \commC_k + (1-\gamma_k) E_k .
\label{eq:optProblET_ideal}
\end{equation}
Ideally, one would like to know $\eI_k$ and $\eII_k$ exactly when computing the estimation cost in order to determine whether it is worth paying the cost for communication. However, $\eI_k$ and $\eII_k$ cannot be computed since the true state is generally unknown (otherwise we would not have to bother with state estimation in the first place).
As is proposed in \cite{TrCa15}, we consider instead the expectation of $E_k$ conditioned on the data $\Dc_k$ that is available by the decision making agent.  Formally, 
\begin{equation}
\min_{\gamma_k \in \{0, 1\}}  \gamma_k \commC_k + (1-\gamma_k) \, \E[ E_k | \Dc_k ] 
\label{eq:optProblET}
\end{equation}
which directly yields the triggering law
\begin{equation}
\text{at time $k$:} \quad \gamma_k = 1  \; \Leftrightarrow \; \E[ E_k | \Dc_k ] \geq C_k .
\label{eq:ETgeneral}
\end{equation}
In \cite{TrCa15}, this framework was used to re-derive common event-triggering mechanisms such as innovation-based triggers \cite{TrDAn11,Tr12,WuJiJoSh13}, or variance-based triggers \cite{TrDAn14b}, depending on whether the current measurement $y_k$ is included in $\Dc_k$, or not. 


\subsection{Predictive triggers}
\label{sec:predTrigger}
This framework can directly be extended to derive a predictive trigger as formulated in \sect \ref{sec:objective}, which makes a communication decision $M$ steps in advance, where $M\!>\!0$ is fixed by the designer.
Hence, we consider the future decision on $\gamma_{k+M}$ and condition the future estimation cost $E_{k+M}$
on $\Dc_k = \Yall_k$, the data available at the current time $k$.  Introducing $\trigsig_{k+M|k} := \E[ E_{k+M} | \Yall_k ]$, the optimization problem \eqref{eq:optProblET_ideal} then becomes
\begin{equation}
\min_{\gamma_{k+M} \in \{0, 1\}}  \gamma_{k+M} \commC_{k+M} + (1-\gamma_{k+M}) \trigsig_{k+M|k} 
\label{eq:optProblPT}
\end{equation}
which yields the \emph{predictive trigger} (PT):
\begin{equation}
\text{at time $k$:} \quad \gamma_{k+M} = 1  \; \Leftrightarrow \; \trigsig_{k+M|k} \geq C_{k+M} .
\label{eq:PTgeneral}
\end{equation}
In \sect \ref{sec:triggers}, we solve $\trigsig_{k+M|k} = \E[ E_{k+M} | \Yall_k ]$ for a specific choice of error measure \eqref{eq:estCostFunction}
to obtain an expression for the trigger \eqref{eq:PTgeneral} in terms of the problem parameters.

\subsection{Self-triggers}
\label{sec:selfTrigger}
A self-trigger computes the next triggering instant at the time when an update is sent.
%
A self triggering law is thus obtained 
 by solving \eqref{eq:PTgeneral} at time $k = \last_k$ for the smallest $M$ such that $\gamma_{k+M} = 1$.  
Recall that $\last_k \leq k$ denotes the last triggering time; in the following, we drop `$k$' when clear from context and simply write $\last_k = \last$.  Formally, the \emph{self-trigger} (ST) is then given by:
\begin{align}
\!\text{at time $k\!=\!\last$:} \,\,\, &\text{find smallest $M\! \geq\! 1$ s.t.\ $\trigsig_{\last+M|\last} \geq C_{\last+M}$}, \nonumber \\[-1mm]
& \text{set} \, \gamma_{\ell+1} \!=\! \dots  \!=\! \gamma_{\ell+M-1}\!=\!0, \gamma_{\ell+M}\!=\!1.
\label{eq:STgeneral}
\end{align}
%

While both the PT and the ST compute the next trigger ahead of time, they represent two different triggering concepts.
The PT \eqref{eq:PTgeneral} is evaluated at every time step $k$ with a fixed prediction horizon $M$, whereas the ST \eqref{eq:STgeneral} needs to be evaluated at $k = \last$ only and yields (potentially varying) $M$.
Which of the two should be used depends on the application (\eg whether continuous monitoring of the error signal is desirable).  In \sect \ref{sec:simulations}, the two concepts are compared in terms of their effectiveness in trading off estimation quality and communication.

\section{Error Distributions} 	
\label{sec:errorDistributions}
In this section, we compute the conditional error PDFs $f(\eI_{k+M} | \Yall_{k})$ and $f(\eII_{k+M} | \Yall_{k})$, which characterize the distribution of the estimation cost $E_{k+M}=E(\eI_{k+M}, \eII_{k+M})$.  These results are used in the next section to solve for the triggers \eqref{eq:PTgeneral} and \eqref{eq:STgeneral} for a specific choice of $E$.


Both triggers \eqref{eq:PTgeneral} and \eqref{eq:STgeneral} predict the communication decisions $M$ steps ahead of the current time $k$ ($M$ is a design parameter for \eqref{eq:PTgeneral} and computed in case of \eqref{eq:STgeneral}).  Hence, in both cases, the set of triggering decisions $\Gamall_{k+M}$ can be computed from the set of measurements $\Yall_k$.  In the following, it will be convenient to denote the index of the last nonzero element in $\Gamall_{k+M}$ (i.e., the last planned triggering instant) by $\lastel_k$; for example, for $\Gamall_{10} = \{ \dots, \gamma_8 = 1, \gamma_9=1, \gamma_{10}=0 \}$, $k=6$, and $M=4$, we have $\lastel_{6} = 9$.  It follows that $\lastel_k \geq \last_k$ in general, with equality $\lastel_k = \last_k$ if no trigger is planned for the next $M$ steps.  

%


The following two lemmas state the sought error PDFs.
\begin{lemma}
\label{lem:PDF_eI}
The predicted error $\eI_{k+M}$ conditioned on $\Yall_k$ is normally distributed, 
\begin{equation}
f(\eI_{k+M} | \Yall_{k}) = 
\Nc(\eI_{k+M}; \, \eIhat_{k+M|k}, \PI_{k+M|k} ) \label{eq:lem1_eIpdf}
\end{equation}
with mean and variance given by, for $k > \lastel_{k-1}$:
\begin{align}
\eIhat_{k+M|k} &= \Phi_{(k+M-1):k} \, (\xKF_k - \xKF_{k|\last} ) 
 \label{eq:lem1_eImean} \\
\PI_{k+M|k} &= \PKF_{k+M|k}  \label{eq:lem1_eIvar} 
\end{align}
and, for $k \leq \lastel_{k-1}$:
\begin{align}
\eIhat_{k+M|k} &= 0  \label{eq:lem1_eImean_b} \\
\PI_{k+M|k} &= \PKF_{\lastel+\Delta|\lastel} = \PKF_{k+M|\lastel}  \label{eq:lem1_eIvar_b}
\end{align}
where 
$\lastel$ is used as shorthand for $\lastel_{k-1}$, and $\Delta := k+M-\kappa$.
\end{lemma}
\begin{lemma}
\label{lem:PDF_eII}
The predicted error $\eII_{k+M}$ conditioned on $\Yall_k$ is normally distributed with
\begin{align}
f(\eII_{k+M} | \Yall_{k}) 
&= \Nc(\eII_{k+M}; \, \eIIhat_{k+M|k}, \PII_{k+M|k} ) \nonumber \\
&= \Nc(\eII_{k+M}; \, 0, \PKF_{k+M} ) . \label{eq:lem2_PDF_eII}
\end{align}
\end{lemma}

We first prove \Lem \ref{lem:PDF_eII}, which will be used in the proof of \Lem \ref{lem:PDF_eI}.

\begin{proof} {\it(Lemma \ref{lem:PDF_eII})}
Because $\xII_k = \xKF_k$ from \eqref{eq:remoteEst},
the error $\eII_k$ is identical to the error $\eKF_k:= x_k - \xKF_k$ of the standard KF \eqref{eq:KF1}--\eqref{eq:KF5}.  From KF theory \cite[p.~41]{AnMo05}, it is known that the conditional and unconditional error distributions are identical, namely
\begin{equation}
f(\eKF_k) = f(\eKF_k | \Yall_k) = \Nc(\eKF_k; 0, \PKF_k) .
\label{eq:lem2_eKF}
\end{equation}
That is, the error distribution is independent of any measurement data.  
%
%
Therefore, we also have $f(\eKF_{k+M} | \Yall_k) = f(\eKF_{k+M})$, which can formally be seen from
\newcommand{\YM}{\Yall_{:M}}
\begin{align}
f(&\eKF_{k+M} | \Yall_k) 
= \int_{\YM} f(\eKF_{k+M} | \YM, \Yall_k)  f(\YM | \Yall_k) \, d\YM \nonumber \\
&= \int_{\YM} \underbrace{f(\eKF_{k+M} | \Yall_{k+M})}_{= f(\eKF_{k+M}) \, \text{(indep.\ of $\YM$)}}  f(\YM | \Yall_k) \, d\YM \nonumber \\
&= f(\eKF_{k+M}) \int_{\YM}  f(\YM | \Yall_k) \, d\YM
= f(\eKF_{k+M})  \label{eq:lem2_fekM}
\end{align}
where $\Yall_{:M}$ denotes the set of measurements $\{y_{k+M}, \dots,$ $y_{k+1} \}$.
The claim then follows from $\eII_{k+M} = \eKF_{k+M}$, \eqref{eq:lem2_fekM}, and \eqref{eq:lem2_eKF}.
\end{proof}

\begin{proof} {\it(Lemma \ref{lem:PDF_eI})}
{\it Case $k > \lastel_{k-1}$:}
First, we note that $k > \lastel_{k-1}$ implies $\lastel_{k-1} = \last_{k}$ because $\lastel_{k-1}$, the last nonzero element of $\Gamall_{k+M-1}$, is in the past, 
and the identity thus follows from the definition of $\last_{k}$.  It follows further that all triggering decisions following $\gamma_\last = 1$ are 0 
until $\gamma_{k+M-1}$ (otherwise $\gamma_\last$ would not be the last element in $\Gamall_{k+M-1}$).  Hence, we have the communication pattern $\gamma_\last = 1$ and $\gamma_{\last+1} = \gamma_{\last+2} = \dots = \gamma_{k+M-1}=0$, and thus from  \eqref{eq:remoteEst},
\newcommand{\myNegSpace}{\!\!\!\!\!\!\!\!\!\!\!\!\!\!\!\!\!\!\!\!\!\!}
\begin{align}
\xI_{k+M} &= A_{k+M-1} \, \hat{x}_{k+M-1} && \myNegSpace \text{(def.\ of $\xI_{k+M}$)} \nonumber \\
&= A_{k+M-1} A_{k+M-2} \, \hat{x}_{k+M-2} && \myNegSpace \text{($\gamma_{k+M-1}=0$)} \nonumber \\
&= \ldots = A_{k+M-1} A_{k+M-2} \cdots A_{\last+1} A_{\last} \, \hat{x}_{\last} \nonumber \\
&= \Phi_{(k+M-1):\last} \, \xKF_{\last} && \myNegSpace \text{($\gamma_{\last}=1$)} . \label{eq:lem1_xhat1}
\end{align}

From \eqref{eq:eI}, it follows
\begin{align}
(\eI_{k+M} | \Yall_k) &= (x_{k+M} | \Yall_k) - \xI_{k+M} . \label{eq:lem1_eI}
\end{align}
where $\xI_{k+M}$ is given by \eqref{eq:lem1_xhat1} and $(x_{k+M} | \Yall_k)$ is Gaussian distributed according to \eqref{eq:PDFstatePredM}. Therefore, $(\eI_{k+M} | \Yall_k)$ is Gaussian with mean $\eIhat_{k+M|k} = \xKF_{k+M|k} - \xI_{k+M}$ and variance $\PI_{k+M|k} = \PKF_{k+M|k}$ \eqref{eq:lem1_eIvar}.
The mean can be rewritten as \eqref{eq:lem1_eImean} using $\Phi_{(k+M-1):\last} = \Phi_{(k+M-1):k} \Phi_{(k-1):\last}$ in \eqref{eq:lem1_xhat1}, and \eqref{eq:KF_meanPred}. This completes the proof for this case.


{\it Case $k \leq \lastel_{k-1}$:}  We use $\lastel = \lastel_{k-1}$ to simplify notation.
By definition of $\lastel$, we have $\lastel \leq M+k-1$, and hence $k \leq \lastel \leq M+k-1$.  That is, a triggering will happen now or 
before the end of the horizon $M+k$.  At the triggering instant $\lastel$, we have from \eqref{eq:remoteEst} and \eqref{eq:eII}, $e_{\lastel} = x_{\lastel} - \xKF_{\lastel}$.
Hence, the distribution of the error at time $\lastel$ is known irrespective of past data $\Yall_k$ and future measurements.  Following the same arguments as in the proof of \Lem \ref{lem:PDF_eII}, we have $f(e_{\lastel} | \Yall_k) = \Nc(e_{\lastel}; \, 0, \PKF_{\lastel})$.  

From the definition of $\lastel$, we know that there is no further communication happening until $M+k-1$.  Thus, we can iterate \eqref{eq:remoteEst} with $\gamma=0$ to obtain the errors $\eI$.
For the first step, we have $\eI_{\lastel + 1} = A_\lastel x_{\lastel} + v_{\lastel} - A_\lastel \xKF_{\lastel} = A_\lastel e_{\lastel} + v_{\lastel}$
and, therefore,  $f(\eI_{\lastel + 1} | \Yall_k) = \Nc(\eI_{\lastel + 1}; \, 0, \Vok{{\lastel}} (\PKF_{\lastel}))$.  Iterating further in the same way until $\lastel + \Delta = k + M$, we obtain \eqref{eq:lem1_eIpdf} with \eqref{eq:lem1_eImean_b} and \eqref{eq:lem1_eIvar_b}.
\end{proof}



\section{Example Triggers} 	
\label{sec:triggers}
Using the triggering framework and results of the previous section, we derive some example triggers next.  For the estimation cost \eqref{eq:estCostFunction}, we specify
\begin{equation}
E_k = (\eI_k)^\transp \eI_k - (\eII_k)^\transp \eII_k .
\label{eq:Ek_squares}
\end{equation}
Other choices for measuring the discrepancy between $\eI$ and $\eII$ are also conceivable, and the framework can be applied analogously. 
The specification \eqref{eq:Ek_squares} is reasonable if keeping the squared estimation error $(e_k)^\transp e_k$ small is of interest, which is a typical objective in estimation.  The estimation cost in \eqref{eq:Ek_squares} is positive if the squared error $(\eI_k)^\transp \eI_k$ (\ie without communication) is larger than $(\eII_k)^\transp \eII_k$ (with communication), which is to be expected on average.  The scalar version \eqref{eq:Ek_squares_scalar} of \eqref{eq:Ek_squares} was used in \cite{TrCa15} to derive optimal event triggers.

\subsection{Self-trigger}
Using the results from the previous section, we solve \eqref{eq:STgeneral} with \eqref{eq:Ek_squares} to obtain a self triggering rule.
Applying \Lem \ref{lem:PDF_eI} (for $k = \last = \lastel_{k-1}$) and \Lem \ref{lem:PDF_eII}, we obtain
\begin{align}
&\trigsig_{\last+M |\last} 
=  \E[ (\eI_{\last+M})^\transp \eI_{\last+M} | \Yall_\last ]  - \E[ (\eII_{\last+M})^\transp \eII_{\last+M} | \Yall_\last ] \nonumber \\
&\phantom{=}= \norm{\eIhat_{\last+M|\last}}^2  - \norm{\eIIhat_{\last+M|\last}}^2 
+ \trace(\PI_{\last+M|\last} - \PII_{\last+M|\last}) \nonumber \\
&\phantom{=}= \trace( \PKF_{\last+M|\last} - \PKF_{\last+M} )
\end{align}
where $\E[e^\transp e] 
= \norm{\E[e]}^2 + \trace(\Var[e])$ with $\norm{\cdot}$ the Euclidean norm was used.
Thus, the self-trigger (ST) \eqref{eq:STgeneral} is: 
\begin{align}
&\text{find smallest $M \geq 1$ s.t.\ $\trace( \PKF_{\last+M|\last} - \PKF_{\last+M} ) \geq C_{\last+M}$}, \nonumber \\[-1mm]
& \text{set} \,\,\, \gamma_{\ell+1} \!=\! \dots  \!=\! \gamma_{\ell+M-1}\!=\!0, \gamma_{\ell+M}\!=\!1.
\label{eq:STsquaredError}
\end{align}
%

%

The self-triggering rule is intuitive: a communication is triggered when the predicted variance $\PKF_{\last+M|\last}$ of the open-loop estimator exceeds the closed-loop variance $\PKF_{\last+M}$ by more than the cost of communication.  The estimation mean does not play a role here, since both open-loop and closed-loop prediction errors $\eI$ and $\eII$ have zero mean for $k = \lastel$.

\subsection{Predictive trigger}
Similarly, we can employ lemmas \ref{lem:PDF_eI} and \ref{lem:PDF_eII} to compute the predictive trigger \eqref{eq:PTgeneral}.  For $k > \lastel_{k-1}$ (\ie the last scheduled trigger occurred in the past), we obtain
\begin{align}
\trigsig_{k+M|k} 
&=  \E[ (\eI_{k+M})^\transp \eI_{k+M} | \Yall_k ]  - \E[ (\eII_{k+M})^\transp \eII_{k+M} | \Yall_k ] \nonumber \\
&= \norm{ \Phi_{(k+M-1):k} (\xKF_k - A_{k-1} \hat{x}_{k-1} ) }^2 \nonumber \\
&\phantom{=} 
+ \trace\big( \PKF_{k+M|k} - \PKF_{k+M} \big)
\label{eq:PTsquaredError1_E}
\end{align}
and, for $k \leq \lastel_{k-1}$ (\ie a trigger is scheduled now or in future), 
\begin{align}
\trigsig_{k+M|k} &= \trace\big(  \PKF_{\lastel + \Delta|\lastel} - \PKF_{\lastel + \Delta} \big) .
\label{eq:PTsquaredError2_E}
\end{align}
In \eqref{eq:PTsquaredError1_E}, we used $\Phi_{(k-1):\last }  \xKF_\last = A_{k-1} \hat{x}_{k-1}$, which follows from the definition of the remote estimator \eqref{eq:remoteEst} with $\gamma_k = 0$ for $k > \last$.
The predictive trigger (PT) \eqref{eq:PTgeneral} is then given, for $k > \lastel_{k-1}$,  by 
\begin{align}
\gamma_{k+M} = 1 \,\, \Leftrightarrow \,\,  
&\norm{ \Phi_{(k+M-1):k} (\xKF_k \! -\! A_{k-1} \hat{x}_{k-1} ) }^2 \nonumber \\
&+ \trace\big( \PKF_{k+M|k} \!-\! \PKF_{k+M} \big) \geq C_{k+M} \label{eq:PTsquaredError1}
\end{align}
and, for $k \leq \lastel_{k-1}$,  by
\begin{align}
&\gamma_{k+M} = 1 \,\, \Leftrightarrow \,\,  
\trace\big( \PKF_{\lastel+\Delta|\lastel} - \PKF_{\lastel + \Delta} \big) \geq C_{\lastel + \Delta}  .
\label{eq:PTsquaredError2} 
\end{align}

%


Similar to the ST \eqref{eq:STsquaredError}, the second term in the PT \eqref{eq:PTsquaredError1} relates the $M$-step open-loop prediction variance $\PKF_{k+M|k}$ to the closed-loop variance $\PKF_{k+M}$.  However, now the reference time is the current time $k$, rather than the last transmission $\last$, because the PT exploits data until $k$.
In contrast to the ST, the PT also includes a mean term (first term in \eqref{eq:PTsquaredError1}).  When conditioning on new measurements $\Yall_k$ ($k>\last$), the remote estimator (which uses only data until $\last$) is biased; that is, the mean of $\eI$ in \eqref{eq:lem1_eImean} is non-zero.  The bias term captures the difference in the mean estimates of the remote estimator ($A_{k-1} \hat{x}_{k-1}$) and the local one ($\xKF_k$), both predicted forward by $M$ steps.
This bias 
contributes to the estimation cost \eqref{eq:PTsquaredError1}.

The rule \eqref{eq:PTsquaredError2} corresponds to the case where a trigger is already scheduled to happen at time  $\lastel$ in future (within the horizon $M$).  Hence, it is clear that the estimation error will be reset at $\lastel$, and from that point onward, variance predictions are used in analogy to the self triggering rule \eqref{eq:STsquaredError} ($\last$ replaced with $\lastel$, and the horizon $M$ with $\Delta$).  This trigger is independent of the data $\Yall_k$ because the error at the future reset time $\lastel$ is fully determined by the distribution \eqref{eq:lem2_PDF_eII}, independent of $\Yall_k$.


\subsection{Discussion}
The derived rules for the ST and the PT have the same threshold structure\footnote{For the ST \eqref{eq:STsquaredError}, \eqref{eq:trigGenStructure} is understood in the sense that \eqref{eq:trigGenStructure} is evaluated for increasing $M\geq 1$ until a positive trigger $\gamma_{k+M}=1$ is found.
} 
\begin{equation}
\gamma_{k+M} = 1 \,\, \Leftrightarrow \,\,  \trigsig_{k+M|k} \geq C_{k+M} 
\label{eq:trigGenStructure}
\end{equation}
where the communication cost $C_{k+M}$ corresponds to the triggering threshold.
The triggers differ in the expected estimation cost $\trigsig_{k+M|k}$. 
 We next analyze the structure of the triggers in more detail.
In addition to the proposed triggers, we also compare to an \emph{event trigger} (ET).
Using the same framework developed herein,
the ET can be obtained from the PT \eqref{eq:PTsquaredError1} by setting $M=0$:
\begin{align}
\gamma_{k} = 1 \,\, \Leftrightarrow \,\, 
\trigsig_{k|k}
&= \norm{ \xKF_k - A_{k-1} \hat{x}_{k-1}  }^2  \label{eq:ETsquaredError} \\
&= \norm{ \xII_k - \xI_k  }^2\geq C_k. \nonumber
\end{align}
The trigger directly compares the two options at the remote estimator, $\xI_k$ and $\xII_k$.
To implement the ET, communication must be available instantaneously if needed.

For the purpose of comparing the structure of the different triggers, we introduce
\begin{align}
\trigsigM_{k,M} &:= \norm{ \Phi_{(k+M-1):k} (\xKF_k \! -\! A_{k-1} \hat{x}_{k-1} ) }^2 \label{eq:trigSigMean} \\
\trigsigV_{k,M} &:= \trace( \PKF_{k+M|k} \!-\! \PKF_{k+M} ). \label{eq:trigSigVar}
\end{align}
The triggers 
ST \eqref{eq:STsquaredError}, PT \eqref{eq:PTsquaredError1}, \eqref{eq:PTsquaredError2}, and ET \eqref{eq:ETsquaredError} can then be characterized as follows.  Each trigger is given by \eqref{eq:trigGenStructure} with
\begin{align}
\trigsig_{k+0|k} &= \trigsigM_{k,0}, M=0 && \text{(ET)} \label{eq:ETcharacterization} \\
\trigsig_{k+M|k} &= \trigsigM_{k,M} + \trigsigV_{k,M} \quad \quad \quad && \text{(PT), $k > \lastel$} \label{eq:PTcharacterization1} \\
\trigsig_{k+M|k} &= \trigsigV_{\kappa,\Delta} && \text{(PT), $k \leq \lastel$} \label{eq:PTcharacterization2} \\
\trigsig_{\last+M|\last} &= \trigsigV_{\last,M} && \text{(ST)} . \label{eq:STcharacterization}
\end{align}
Hence, the trigger signals are generally a combination of the `mean' signal \eqref{eq:trigSigMean} and the `variance' signal \eqref{eq:trigSigVar}.  Noting that the mean signal \eqref{eq:trigSigMean} depends on real-time measurement data $\Yall_k$ (through $\xKF_k$), while the variance signal \eqref{eq:trigSigVar} does not, we can characterize ET and PT as 
\emph{online triggers}, while ST is an 
\emph{offline trigger}.  This reflects the intended design of the different triggers. ST is designed to predict the next trigger at the time $\last$ of the last triggering, without seeing any data beyond $\last$.  This allows the sensor to go sleep in-between triggers, for example.  ET and PT, on the other hand, continuously monitor the sensor data to make more informed transmit decisions (as shall be seen when comparing the effectiveness of the different triggers in \sect \ref{sec:simulations}).  

While ET requires instantaneous communication, 
which is limiting for online allocation of communication resources,
PT makes the transmit decision $M\geq1$ steps ahead of time.  ET compares the mean estimates only (\cf \eqref{eq:ETcharacterization}), while PT results in a combination of mean and variance signal (\cf \eqref{eq:PTcharacterization1}).  If a transmission is already scheduled for $\lastel_{k-1} \geq k$, PT resorts to the ST mechanism for predicting beyond $\lastel_{k-1}$; that is, it relies on the variance signal only (\cf \eqref{eq:PTcharacterization2}).
  
While ST can be understood as an \emph{open-loop} trigger (\eqref{eq:STcharacterization} can be computed without any measurement data), ET clearly is a \emph{closed-loop} trigger requiring real-time data $\Yall_k$ for the decision on $\gamma_k$.  PT can be regarded as an intermediate scheme exploiting real-time data and variance-based predictions.  Accordingly, the novel predictive triggering concept lies between the known concepts of event and self triggering.

The ST is similar to the variance-based triggers proposed in \cite{TrDAn14b}.  
Therein, it was shown for a slightly different scenario (transmission of measurements instead of estimates) that event triggering decisions based on the variance are independent of any measurement data and can hence be computed off-line.
Similarly, when assuming that all problem parameters $A_k$, $H_k$, $Q_k$, $R_k$ in \eqref{eq:sys_x}, \eqref{eq:sys_y} are known a-priori, 
\eqref{eq:STsquaredError} can be pre-computed for all times.  However, if some parameters only become available during operation (\eg the sensor accuracy $R_k$), the ST also becomes an online trigger. 

\section{Illustrative Numerical Examples} 	
\label{sec:simulations}
Key characteristics of 
the proposed self-trigger (ST) and predictive trigger (PT) are illustrated through numerical simulations of stable and unstable scalar processes.

\subsection{Behavior of self-trigger}
\begin{figure}[tb]
\centering
\includegraphics[scale=.9]{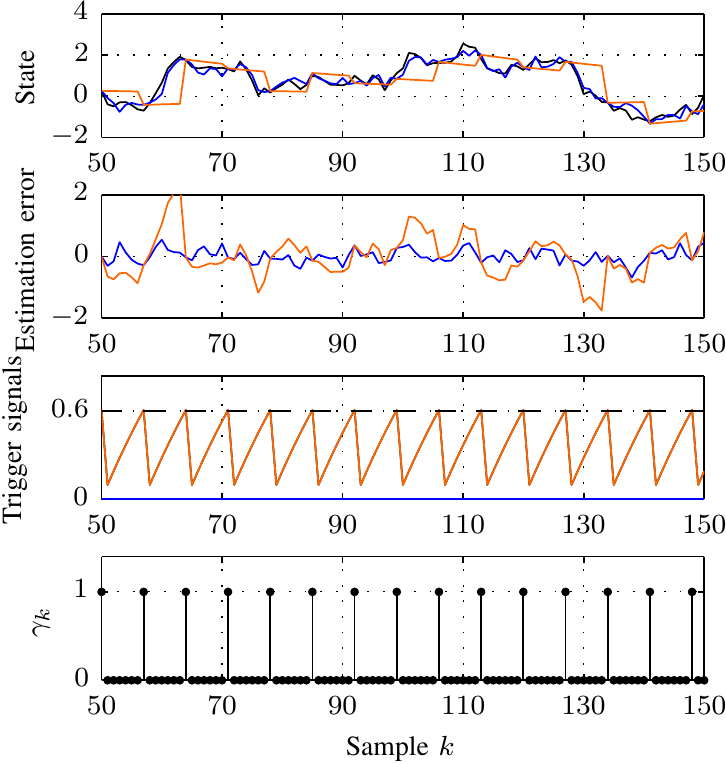}	
\caption{Example~\ref{ex:ex1} with self-trigger (ST).  The TOP graph shows the true state $x$ (\graph{black}), the KF estimate $\xKF$ (\graph{blue}), and the remote estimate $\hat{x}$ (\graph{orange}); and in the SECOND graph are the corresponding errors $\eKF=x-\xKF$ (\graph{blue}) and $e=x-\hat{x}$ (\graph{orange}).  The THIRD graph shows 
$\trigsigM$ \eqref{eq:trigSigMean} (\graph{blue}), $\trigsigV$ \eqref{eq:trigSigVar} (\graph{orange}), the triggering signal $\trigsig = \trigsigM + \trigsigV$ (\graph{black}, hidden), and the threshold $C_k = 0.6$ (\graph{dashed}).  The BOTTOM graph indicates the triggering decisions $\gamma$.
}
\label{fig:example1_1}
\end{figure}

First, we consider simulations of the stable, scalar, linear time-invariant (LTI) process:
\begin{example}
$A_k = 0.98$, $H_k = 1$, $Q_k = 0.1$, $R_k = 0.1$ for all $k$, and $\bar{x}_0 = X_0 = 1$.
\label{ex:ex1}
\end{example}
Results of the numerical simulation of the event-based estimation system consisting of (\cf \fig \ref{fig:remoteEstimation}) the local estimator \eqref{eq:KF1}--\eqref{eq:KF5}, the remote estimator \eqref{eq:remoteEst}, and the ST \eqref{eq:STsquaredError} with constant cost $C_k = C= 0.6$ are shown in \fig \ref{fig:example1_1}.
The estimation errors of the local and remote estimator are compared in the second graph.  
As expected, the remote estimation error $e_k = x_k-\hat{x}_k$ (orange) is larger than the local estimation error $\eKF_k = x_k-\xKF_k$ (blue).  Yet, the remote estimator only needs 14\% of the samples.  

The triggering behavior is illustrated in the third graph showing the triggering signals $\trigsigM$ \eqref{eq:trigSigMean}, $\trigsigV$ \eqref{eq:trigSigVar}, and $\trigsig = \trigsigM + \trigsigV$, and the bottom graph depicting the triggering decision $\gamma$.
Obviously, the ST entirely depends on the variance signal $\trigsigV$ (orange, identical with $\trigsig$ in black), while $\trigsigM = 0$ (blue).  This reflects the previous discussion about the ST being independent of online measurement data.  
The triggering behavior (the signal $\trigsig$ and the decisions $\gamma$) is actually  \emph{periodic}, which can be deduced as follows:
the variance $\PKF_k$ of the KF \eqref{eq:KF1}--\eqref{eq:KF5} converges exponentially to a steady-state solution $\PKFss$, \cite{AnMo05}; hence, the triggering law \eqref{eq:STsquaredError} asymptotically becomes 
\begin{align}
\trace( \Vo^M(\PKFss) - \PKFss ) \geq C, \quad  \Vo(X) := AXA^\transp + Q
\label{eq:STsquaredErrorLTI_ex1}
\end{align}
and \eqref{eq:STgeneral} thus has a unique (time-invariant) solution $M$ corresponding to the period seen in \fig \ref{fig:example1_1}.  

Periodic transmit sequences are typical for variance-based triggering on time-invariant problems, which has also been found and formally proven for related scenarios in \cite{TrDAn14b,LeDeQu15}.  

\subsection{Behavior of predictive trigger}
\begin{figure}[tb]
\centering
\includegraphics[scale=.9]{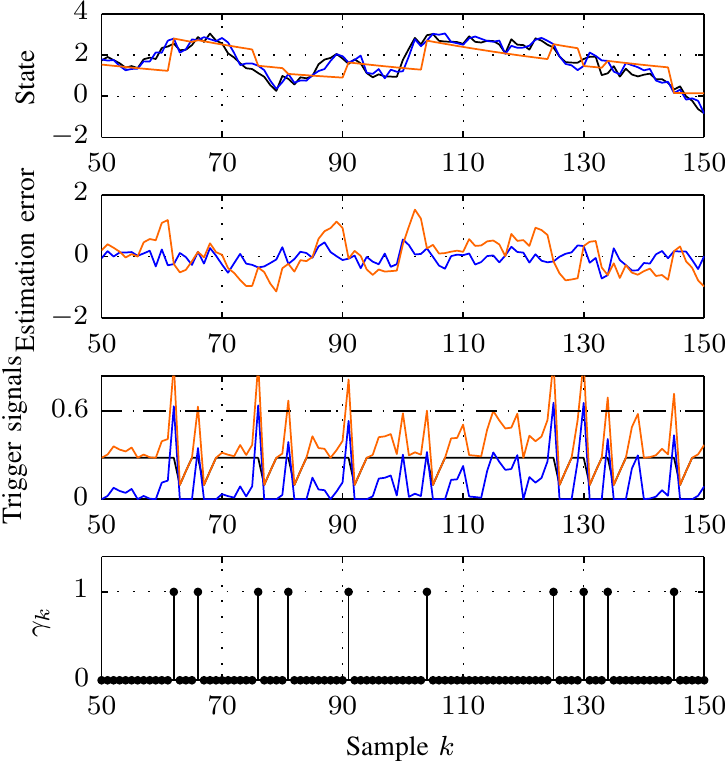}
\caption{Example~\ref{ex:ex1} with predictive trigger (PT) and $C_k = 0.6$.  Coloring of the signals is the same as in \fig \ref{fig:example1_1}.  The triggering behavior is \emph{stochastic}.
}
\label{fig:example1_2}
\end{figure}

\begin{figure}[tb]
\centering
\includegraphics[scale=.9]{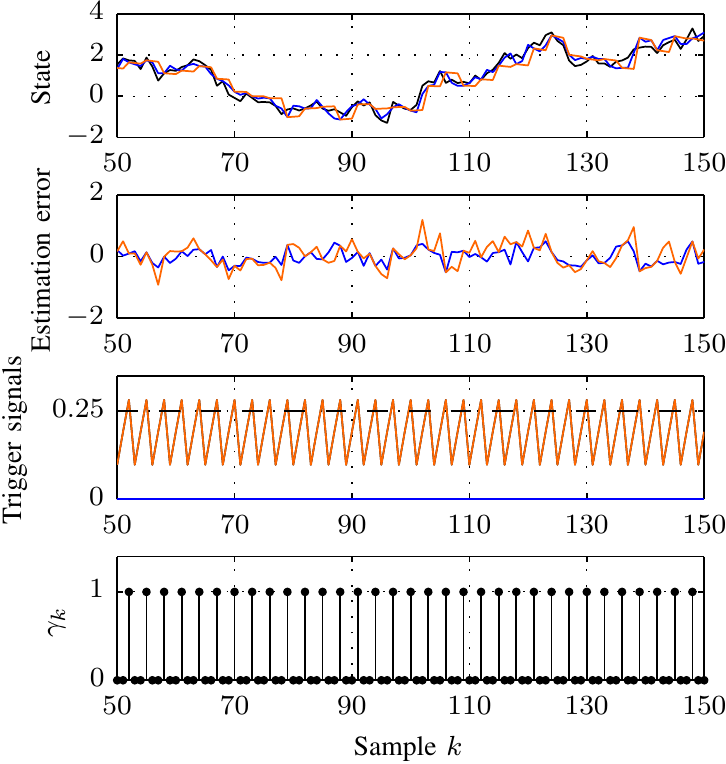}
\caption{Example~\ref{ex:ex1} with predictive trigger (PT) and $C_k = 0.25$.  Coloring of the signals is the same as in \fig \ref{fig:example1_1}. The triggering behavior is \emph{periodic}.
}
\label{fig:example1_3}
\end{figure}

The results of simulating \Ex \ref{ex:ex1}, now with 
the PT \eqref{eq:PTsquaredError1}, \eqref{eq:PTsquaredError2}, and prediction horizon $M=2$, are presented in \fig \ref{fig:example1_2} for the cost $C_k = C = 0.6$, and in \fig \ref{fig:example1_3} for $C_k = C = 0.25$.  Albeit using the same trigger, the two simulations show fundamentally different triggering behavior: while the triggering signal $\trigsig$ and the decisions $\gamma$ in \fig \ref{fig:example1_2} are irregular, they are periodic in \fig \ref{fig:example1_3}.  

Apparently, the choice of the cost $C_k$ determines the different behavior of the PT.
For $C_k = 0.6$, the triggering decision depends on both, the mean signal $\trigsigM$ and the variance signal $\trigsigV$, as can be seen from \fig \ref{fig:example1_2} (third graph).
Because $\trigsigM$ is based on real-time measurements,
which are themselves random variables \eqref{eq:sys_y}, the triggering decision is a random variable.  
We also observe in \fig \ref{fig:example1_2} that the variance signal $\trigsigV$ is alone not sufficient to trigger  a communication.\footnote{After convergence of the local estimator variance $\PKF_k$, $\trigsigV$ corresponds to \eqref{eq:STsquaredErrorLTI_ex1}, which does not exceed the chosen $C_k$ for $M=2$ iterations.}
However, when lowering the cost of communication $C_k$ enough, the variance signal alone becomes sufficient to cause triggers.  Essentially, triggering then happens 
according to \eqref{eq:PTsquaredError2} only, and  \eqref{eq:PTsquaredError1} becomes irrelevant.  Hence, the PT resorts to self triggering behavior for small enough communication cost $C_k$.  That is, the PT undergoes a phase transition for some value of $C_k$ from stochastic/online triggering to deterministic/offline triggering behavior.


\subsection{Estimation versus communication trade-off}
Following the same approach as in \cite{TrCa15}, we evaluate the effectiveness of different triggers by comparing their trade-off curves of average estimation error $\Ec$ versus average communication $\Cc$ obtained from Monte Carlo simulations.  In addition to the ST \eqref{eq:STsquaredError} and the PT \eqref{eq:PTsquaredError1}, \eqref{eq:PTsquaredError2}, $M=2$, we also compare against the ET \eqref{eq:ETsquaredError}.  The latter is expected to yield the best trade-off because it makes the triggering decision at the latest possible time (ET decides at time $k$ about communication at time $k$).  

\begin{figure}[tb]
\centering
\includegraphics[scale=0.98]{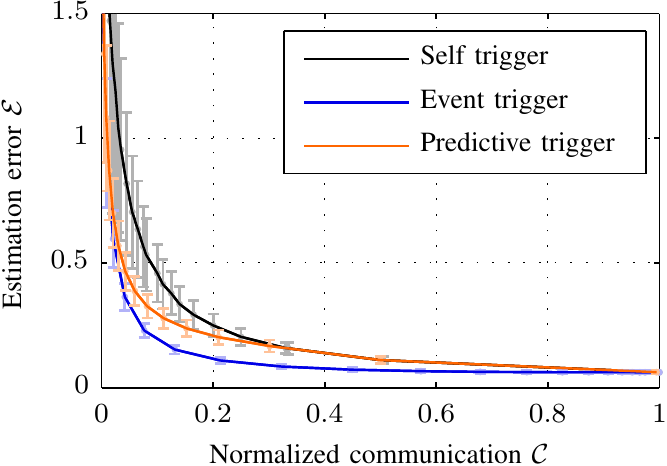} 
\caption{Trade-off between estimation error $\Ec$ and average communication $\Cc$ for different triggering concepts applied to \Ex \ref{ex:ex1} (stable process). 
Each point represents the average from 50'000 Monte Carlo simulations, and the light error bars correspond to one standard deviation. It can be seen that the novel concept of predictive triggering provides a middle ground between event triggering and self triggering. }
\label{fig:example1_EvsC}
\end{figure}

The estimation error $\Ec$ is measured as the squared error $e_k^2$ averaged over the simulation horizon (here, 200 samples) and 50'000 simulation runs.  The average communication $\Cc$ is normalized such that $\Cc=1$ means $\gamma_k=1$ for all $k$, and $\Cc=0$ means no communication (except for one enforced initial communication at $k=1$).  
By varying the constant communication cost $C_k = C$ in a suitable range, an $\Ec$-vs-$\Cc$ curve is obtained, which represents the estimation/communication trade-off for a particular trigger.  The results 
for \Ex \ref{ex:ex1} are presented in \fig \ref{fig:example1_EvsC}.


Comparing the three different triggering schemes, we see that the ET is superior, as expected, because its curve is uniformly below the others.  Also expected, the ST is the least effective 
since no real-time information is available and triggers are purely based on variance predictions.  
The novel concept of predictive triggering can be understood as an intermediate solution between these two extremes.  For small communication cost $C_k$ (and thus relatively large communication $\Cc$), the PT behaves like the ST, as was discussed in the previous section and is confirmed in \fig \ref{fig:example1_EvsC} (orange and black curves essentially identical for large $\Cc$).
When the triggering threshold $C_k$ is relaxed (\ie the cost increased), the PT also exploits real-time data for the triggering decision (through \eqref{eq:trigSigMean}), similar to the ET.  Yet, the PT must predict the decision $M$ steps in advance making its $\Ec$-vs-$\Cc$ trade-off generally less effective than the ET.  In \fig \ref{fig:example1_EvsC}, the curve for PT is thus between ET and ST and approaches either one of them for small and large communication $\Cc$.

\begin{figure}[tb]
\centering
\includegraphics[scale=0.98]{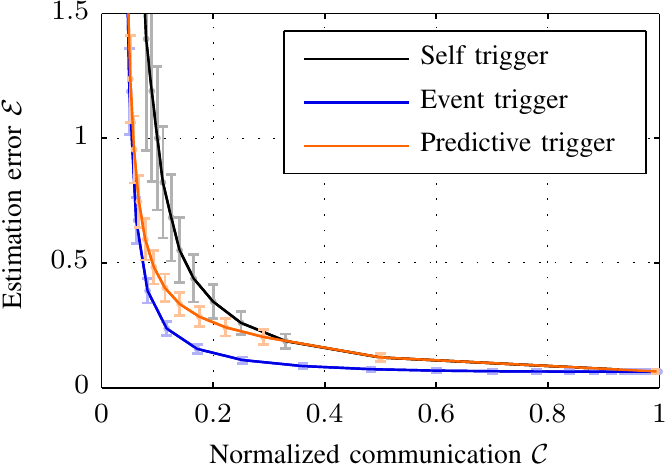} 
\caption{Estimation-vs-communication trade-offs for investigated triggering concepts applied to \Ex \ref{ex:ex2} (unstable process).
}
\label{fig:example2_EvsC}
\end{figure}

Fig.\ \ref{fig:example2_EvsC} shows the $\Ec$-vs-$\Cc$ curves for the unstable system:
\begin{example}
$A_k = 1.1$, $H_k = 1$, $Q_k = 0.1$, $R_k = 0.1$ for all $k$, and $\bar{x}_0 = X_0 = 1$.
\label{ex:ex2}
\end{example}
The same qualitative behavior of the different triggering mechanisms as in \fig \ref{fig:example1_EvsC} can be observed.
%


\section{Concluding Remarks} 	
\label{sec:conclusion}
For event-triggered control and estimation methods to be adopted in industrial practice, it is important to demonstrate tangible benefits in terms of reduced hardware costs or better resource utilization compared to traditional designs.  To achieve, for instance, better utilization of shared network bandwidth, we believe that the tight integration of the control system and the communication system is critical.  In particular, the control system must signal to the communication system its needs for communication ahead of time in order to give the communication system time to respond and reconfigure accordingly.
In this paper, we developed a general framework for making predictive triggering decisions in state estimation, from which we derived two particular types of triggers. 

With the self-trigger (ST), the next triggering instant is computed at the time of data transmission.  The next triggering instant can thus be encoded in the transmitted data packet and, for example, be used by a network manager to reconfigure the network for the next communication round. 
In contrast to the ST, the predictive trigger (PT) continuously reads sensor values and predicts whether communication is needed at $M > 0$ steps in future. The horizon $M$ can be chosen to allow enough time for the communication system to respond to communication requests.
Predictive triggering is a new  concept in-between the known concepts of self triggering and event triggering for estimation, as is shown in the analysis and simulation results herein.

This paper focuses on the fundamental trigger design and, to this end, considers the basic 
remote estimation problem in \fig \ref{fig:remoteEstimation} with a single triggering agent.  Ultimately, we aim at extending these ideas to distributed systems with multiple agents connected over wireless networks.  In particular, we intend to extend and combine prior work on distributed event-based estimation \cite{TrDAn11,Tr12} with recent methods for efficient and reliable communication over multi-hop low-power wireless networks \cite{FeZiMoTh12}.

\bibliographystyle{IEEEtran}
\bibliography{SebDatabase}

\end{document}